\documentclass[journal]{IEEEtran}
\IEEEoverridecommandlockouts

\usepackage{cite}
\usepackage{amsmath,amssymb,amsfonts,amsthm,mathtools}

\DeclarePairedDelimiter{\ceil}{\lceil}{\rceil}

\usepackage{graphicx}
\usepackage{textcomp}
\usepackage{xcolor}
\usepackage{cases}
\usepackage[utf8]{inputenc}
\def\BibTeX{{\rm B\kern-.05em{\sc i\kern-.025em b}\kern-.08em
    T\kern-.1667em\lower.7ex\hbox{E}\kern-.125emX}}
    
\usepackage{epstopdf}

\usepackage{dirtytalk}

\usepackage{enumitem}

\usepackage{algpseudocode}
\usepackage[linesnumbered,ruled,vlined]{algorithm2e}

\usepackage{ragged2e}

\usepackage{multirow}
\usepackage{dcolumn,booktabs}

\usepackage{subcaption}

\usepackage{empheq}

\begin{document}

\title{Vision-Assisted User Clustering for Robust \\  mmWave-NOMA Systems 
}

\author{
    \IEEEauthorblockN{Aditya S. Rajasekaran\IEEEauthorrefmark{2}\IEEEauthorrefmark{3}, Hamza U. Sokun\IEEEauthorrefmark{3}, Omar~Maraqa\IEEEauthorrefmark{1}, Halim Yanikomeroglu\IEEEauthorrefmark{2}, Saad Al-Ahmadi\IEEEauthorrefmark{1}}
    \IEEEauthorblockA{\IEEEauthorrefmark{3}Ericsson Canada Inc, Ottawa, ON K2K 2V6, Canada}
    \IEEEauthorblockA{\IEEEauthorrefmark{2}Department of Systems and Computer Engineering, Carleton University, Ottawa, ON K1S 5B6, Canada}
    \IEEEauthorblockA{\IEEEauthorrefmark{1}Department of Electrical Engineering, and Center of Communication Systems and Sensing,\\ King Fahd University of Petroleum \& Minerals, Dhahran-31261, Saudi Arabia}
    
\thanks{$\copyright$ 2021 IEEE. Personal use of this material is permitted. Permission from IEEE must be obtained for all other uses, in any current or future media, including reprinting/republishing this material for advertising or promotional purposes, creating new collective works, for resale or redistribution to servers or lists, or reuse of any copyrighted component of this work in other works.}%

}

\markboth{Accepted in the Proceedings of IEEE Future Networks World Forum (IEEE FNWF 2022)}
{Rajasekaran \MakeLowercase{\textit{et al.}}: Vision-Assisted User Clustering for Robust mmWave-NOMA Systems}

\maketitle

\begin{abstract}
When operated in the mmWave band, user channels get highly correlated which can be exploited in mmWave-NOMA systems to cluster a set of ``correlated" users together. 
Identifying the set of users to cluster greatly affects the viability of NOMA systems. Typically, only channel state information (CSI) is used to make these clustering decisions. When any problem arises in accessing up-to-date and accurate CSI, user clustering will not properly function due to its hard-dependency on CSI, and obviously, this will negatively affect the robustness of the NOMA systems. To improve the robustness of the NOMA systems, we propose to utilize
emerging trends such as location-aware and camera-equipped base stations (CBSs) which do not require any extra radio frequency resource consumption. Specifically, we explore three different dimensions of feedback that a CBS can benefit from to solve the user clustering problem, namely CSI-based feedback and non-CSI-based feedback, comprised of user equipment (UE) location and the CBS camera feed. We first investigate how the vision assistance of a CBS can be used in conjunction with other dimensions of feedback to make clustering decisions in various scenarios.
Later, we provide a simple user case study to illustrate how to implement vision-assisted user clustering in mmWave-NOMA systems to improve robustness, in which a deep learning (DL) beam selection algorithm is trained on the images captured by the CBS to perform NOMA clustering.
We demonstrate that user clustering without CSI can achieve comparable performance to accurate CSI-based solutions, and user clustering can continue to function without much performance loss even in the scenarios where CSI is severely outdated or not available at all.
\end{abstract}


\begin{IEEEkeywords}
Non-orthogonal multiple access (NOMA); User Clustering; Beamforming (BF); Camera Base Station (CBS); Deep Learning (DL)
\end{IEEEkeywords}

\section{Introduction}


Non-orthogonal multiple access (NOMA) techniques offer a way to serve multiple users in the same orthogonal resource (e.g., time, frequency, orthogonal frequency division multiplexing resource block (RB), etc.) by separating the users in the power or code domains instead. In mmWave bands, users' channels are strongly correlated due to the highly directional nature of mmWave transmission~\cite{xiao2018Joint, Ding2017BF_mmWaveNOMA}. The strong correlations among users' channels in mmWave and higher bands make them ideal for the formation of user clusters that can be served by a single beam and separated in the power or code domain through NOMA. 

\textcolor{black}{For mmWave-NOMA, the way how to cluster users is one of the important functionalities to achieve the desired level of network performance. Obtaining the global solution for that problem, particularly for large-size networks, could be a formidable task due to its combinatorial nature, but local solutions could be efficiently obtained, for instance, using optimization schemes (e.g.,~\cite{Ding2017BF_mmWaveNOMA, rajasekaran_userclustering_2020}) or machine learning (ML) approaches~\cite{cui2018unsupervised, Marasinghe2020AHC}. However, these approaches are all based on the strong assumption of accurate instantaneous channel state information (CSI) from the users. Firstly, since clustering is an aspect of user scheduling, executing clustering algorithms based on instantaneous CSI is problematic, as the CSI acquired might be stale by the time it is used for user clustering. Secondly, for mmWave-NOMA systems, CSI acquisition and user tracking, which are required to establish and support highly directional transmission links, can create a tremendous amount of overhead and latency~\cite{CSI_NR}. Thirdly, in practice, the availability and the reliability of CSI at base stations cannot be always guaranteed due to the following reasons: 1) Non-ideal hardware behavior. 2) Poor performance of physical downlink control channel. 3) Poor performance of physical uplink shared channel. 4) Errors in reported CSI. 5) Errors in decoded CSI. 6) Long CSI reporting duration and having outdated CSI. 7) Frequency gap between uplink and downlink in frequency division duplex mode. All these make it important to find other dimensions of user feedback that the BS can exploit for NOMA clustering decisions, such as user location information or pictures from a scene captured by a camera-equipped BS~(CBS).}

Beyond 5G (B5G) systems can access the location information of users~\cite{magazine2014_location_5g} and exploit the directional nature of mmWave transmission, location aided beamforming (BF) strategies have been developed, e.g.,~\cite{garcia2016_location_mmwave}. Extending this idea to NOMA systems, in~\cite{orikhumi2020_location_clustering_noma_mmwave}, a location-aided NOMA clustering strategy was developed that exploits the user location to assign a user to pre-defined cluster angles. CBSs, on the other hand, can be used to capture red-green-blue (RGB) images of users at a scene and utilize the vast potential of deep learning (DL) algorithms on these images for performing wireless communication tasks~\cite{alrabeiah_viwi_2020, ying_vision_2020, alrabeiah_beamprediction_2020}. In~\cite{alrabeiah_viwi_2020}, a synthetic data generation framework for RGB images and the associated user channels for mobile users was developed. In~\cite{alrabeiah_beamprediction_2020}, for instance, Alrabeiah~\textit{et al.} applied convolutional neural networks (CNN) to residual networks (ResNets)~\cite{he_Resnet_2016} to solve a beam prediction problem. An 18-layer residual network (ResNet-18) was adopted and customized to fit the beam prediction problem.

In this paper, we motivate the use of CBSs \textcolor{black}{to enhance the performance and robustness of mmWave-NOMA systems, which does not need to consume any extra radio frequency (RF) resource. The camera feed of CBSs can be used for user clustering exclusively when CSI is hard to obtain, or the feed can be used in conjunction with CSI not only to improve the accuracy and the quality of the scheduling decisions, but also to reduce the amount of overhead and power consumption in the system.} We highlight the different dimensions of user equipment (UE) feedback that a CBS can exploit for NOMA clustering, namely CSI-based feedback from the user and non-CSI-based feedback, comprised of the images captured by the CBS and UE location. Through a simple case study that performs NOMA clustering exclusively based on CBS images and user location feedback, we show that applying deep learning techniques to the images captured by the CBS can be exploited to achieve a spectral efficiency performance comparable to where NOMA clustering decisions are taken using the full accurate CSI of users. The results of our investigation highlight that in practical NOMA systems, either the CSI of users or visual feed of cameras or some combination of the two can be used interchangeably for NOMA clustering, depending on what feedback is available to the CBS in different situations.

\section{Why camera-equipped BS in mmWave-NOMA clustering?}\label{sec:whyCamera}

\subsection{Dimension space for possible clustering approaches in mmWave-NOMA systems}\label{sec:3dimesnions}


\begin{figure*}[ht!]
  \centering
  \includegraphics[width=0.62\linewidth]{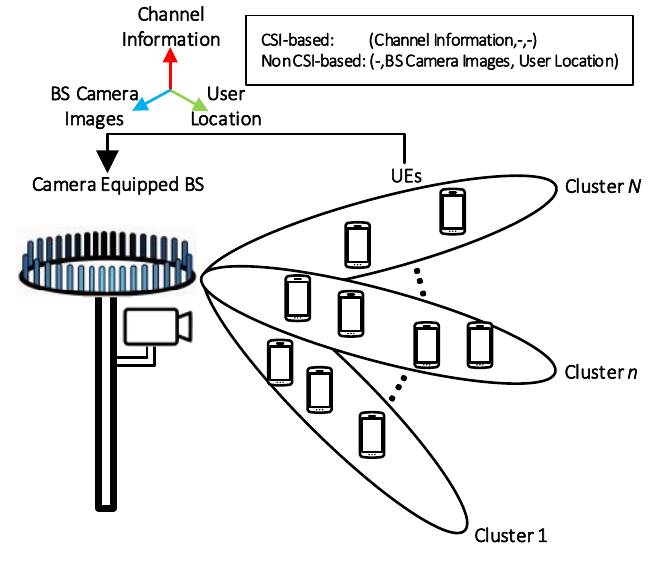}
  \caption{Illustrating the dimensions of feedback that a CBS can exploit for NOMA clustering decisions, categorized into CSI-based feedback and non-CSI-based feedback, comprised of UE location and BS camera feed images.}
  \label{fig:3dimensions}
\vspace{-1.0em}
\end{figure*}

\begin{figure*}[ht!]
  \centering
  \includegraphics[width=0.8\linewidth]{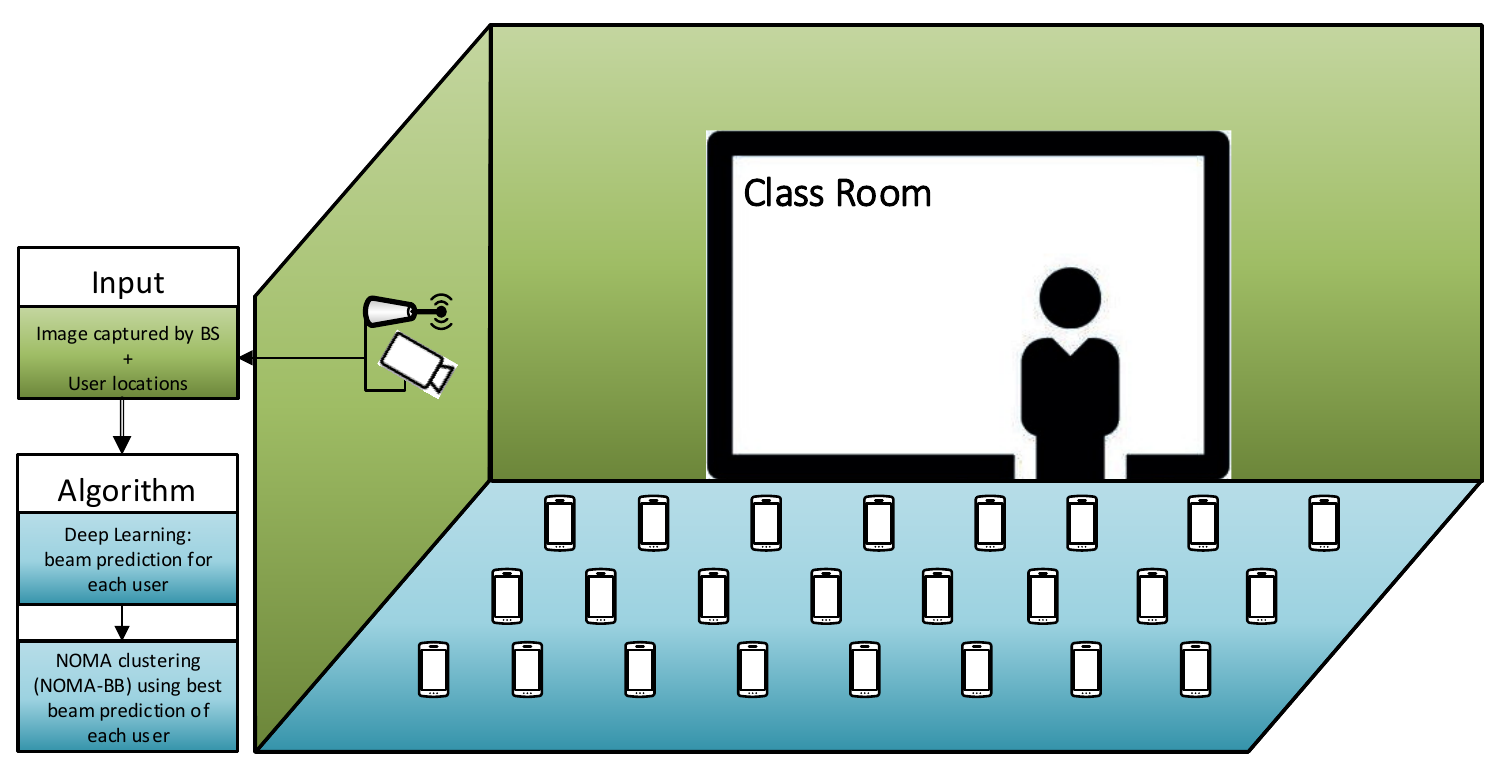}
  \caption{High-level illustration of the proposed NOMA clustering scheme based on non-CSI-based feedback involving the images captured by the camera feed of the BS and UE location input.}
  \label{fig:ClassRoom}
\vspace{-1.5em}
\end{figure*}

In this section, we illustrate the NOMA clustering problem statement for a CBS and the different dimensions of UE feedback that a CBS can use to solve this problem. Consider a multi-antenna, mmWave-NOMA enabled BS as shown in Fig.~\ref{fig:3dimensions}. Using the spatial dimension, users can be separated through beamforming. As we can see in Fig.~\ref{fig:3dimensions}, NOMA allows multiple users to be served in one beam. The goal of user clustering schemes in mmWave-NOMA systems is to identify sets of correlated users, so that they can be grouped into a NOMA cluster. Within a cluster, the users are separated via the power domain, called power-domain NOMA~\cite{maraqa_nomasurvey2020}. Clusters are separated from each other in the spatial domain, using beamforming techniques. Clustering, therefore, represents a user scheduling problem of identifying the set of users to be served in each time slot and of identifying the candidate beams, such that a minimum quality-of-service is guaranteed to each user.

To solve this NOMA clustering problem, we can see that Fig.~\ref{fig:3dimensions} shows the different dimensions of user feedback that a CBS can exploit. This feedback that the CBS relies on for NOMA clustering can be classified under two categories: CSI-based feedback and non-CSI-based feedback. The CSI-based feedback (i.e., the channel information dimension shown in the figure) incurs wireless channel overhead. The non-CSI-based feedback is comprised of two entities, namely user location feedback and pictures from the camera feed of the CBS. The latter is the main focus of this article. Any combination of these three dimensions of feedback may be available at different times, for different users, and in different situations. We note that while obtaining BS images and user location has an associated cost, it is not a cost on the wireless channel resources itself, the most precious commodity in a wireless system. With ever-decreasing camera prices and location information already a common feature of 5G systems~\cite{magazine2014_location_5g}, it is not unrealistic to expect that CBSs will be able to exploit this information in the near future. CBSs can make clustering decisions with information from anyone dimension or combination of dimensions, depending on what is available. 

\subsection{Role of the vision dimension in clustering}

To make clustering decisions, current mmWave-NOMA clustering schemes are based on the CSI feedback from the users, typically using the cosine similarity or Euclidean distance metrics to determine the correlation between users~\cite{cui2018unsupervised}.
A CBS allows for a new dimension of UE feedback that can be used for UE clustering decisions in NOMA systems. The visual dimension can be used in conjunction with or as a replacement for CSI, which is traditionally used for NOMA clustering. In other words, the visual information provided by an external unit can be used by the baseband on the CBS to make a decision about how to form the NOMA clusters using the following possible permutations: 
\begin{itemize}
    \item In case there is no access to the visual information, the CBS will rely on the channel information.
    \item In case there is no access to the channel information, the CBS will rely on the visual information.
    \item In case there is access to both channel and visual information, the CBS can combine information from both sources, \textcolor{black}{e.g., through weighted summation,} to make a better decision. Or, the CBS can resort to visual information as a fallback solution, only when the reliability of channel information is low. To detect whether there is a reliability issue with channel information, the number of consecutive Hybrid Automatic Repeat Request (HARQ) Non-Acknowledgements (NACKs) and the number of consecutive Discontinuous Transmissions (DTXs) might be used.
\end{itemize}

Obtaining CSI is often challenging and problematic in terms of its availability and reliability~\cite{CSI_NR}. This is particularly true for NOMA systems that are designed to serve a large number of users. Even when CSI is available, there is a large cost involved in obtaining the channel information of all users; something that conventional NOMA clustering schemes rely on to work. In particular, obtaining CSI for clustering in this way is wasteful for users who cannot be scheduled on the basis of current CSI, and a newly updated CSI is needed when they are actually scheduled. Additionally, using instantaneous CSI for UE scheduling decisions in future transmission slots presents issues related to the quality of CSI being used. Hence, even when CSI is available, there is an incentive to rely on CSI-free NOMA clustering approaches.

In mmWave systems that have dominant line-of-sight (LoS) paths, approaches such as the location-aware NOMA clustering scheme proposed in~\cite{orikhumi2020_location_clustering_noma_mmwave} can be used. However, a NOMA clustering approach based exclusively on user location feedback is limited to LoS paths and simple channel settings, such that the best beam for a user can be determined exclusively on the basis of user location and no other feature of the channel or surroundings. However, with CBSs, we can feed captured images to powerful DL image processing techniques. These DL techniques, using neural networks, can learn advanced features of the channel and make good beam predictions or cluster formation decisions. These neural networks can also be fed the location of the users as an additional input, particularly for identifying users in images captured by the CBS. Fig.~\ref{fig:ClassRoom} illustrates an example with a classroom deployment where a CBS captures the image of all users along with user location feedback. A deep learning algorithm parses the images from the classroom, using the user location feedback to identify the users. From a codebook of beams, the best beam for each user can be predicted; which can in turn be used for the clustering of users that all select the same best beam. This concept is illustrated through a case study in the next section.

\section{A Case Study: Robust User Clustering in mmWave-NOMA via Vision-Based Deep Learning}

\begin{figure}[t!]
  \centering
  \includegraphics[width=0.8\linewidth]{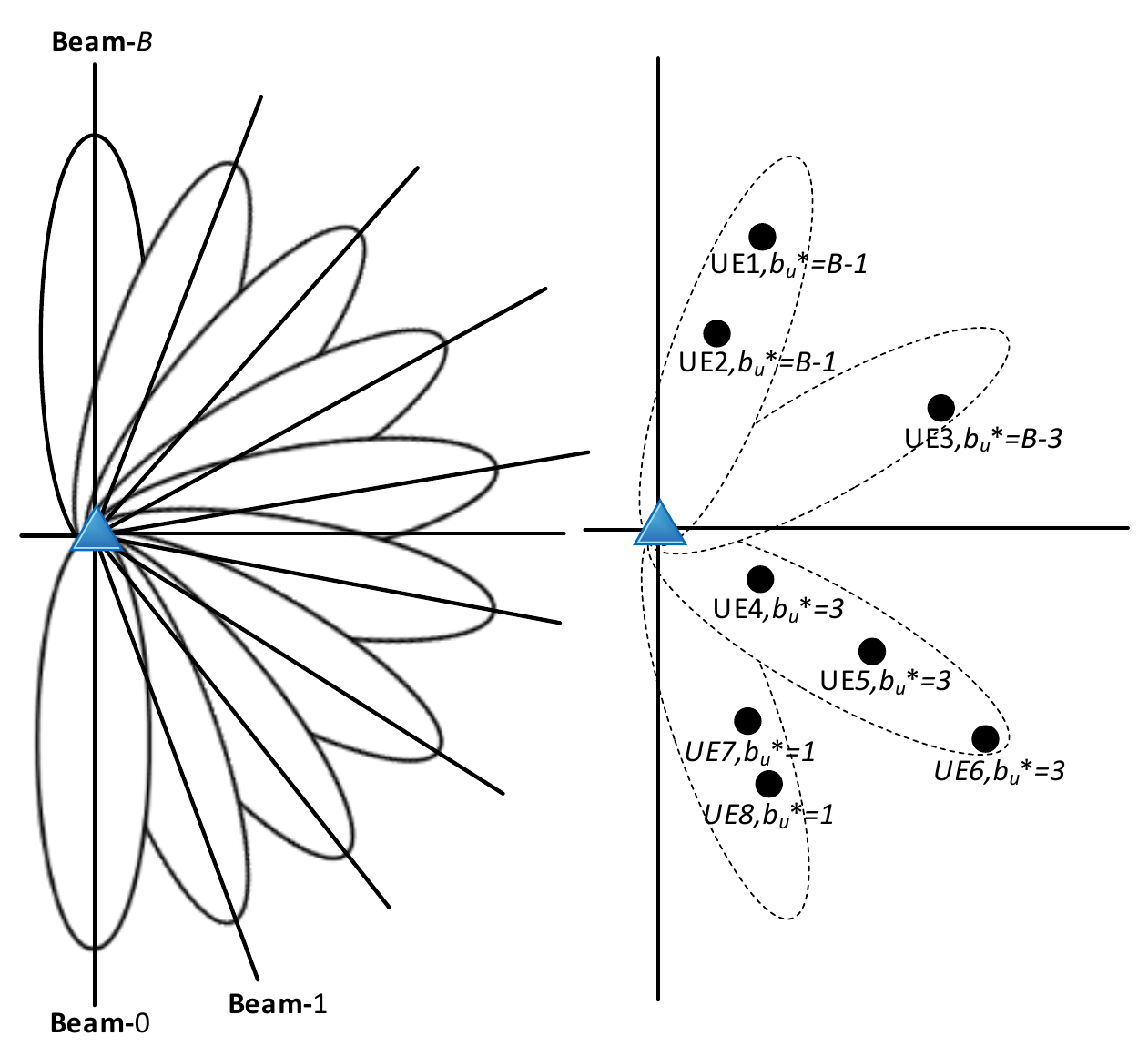}
  \caption{Illustrating the NOMA-BB algorithm, where from a codebook of beams each users best beam is determined and users that share the same best beam are clustered together.}
  \label{fig:systemModel}
\vspace{-1.5em}
\end{figure}


We start by describing a simple heuristic CSI-based mmWave-NOMA clustering scheme, called the NOMA Best Beam (NOMA-BB) clustering approach. We focus on a single-path mmWave channel model in this case study. From a codebook of beams, such as the one shown in Fig.~\ref{fig:systemModel}, the idea behind NOMA-BB is for the BS to identify the best beam for each user from the set of candidate beams. Using the available CSI of each user, the BS uses the cosine similarity metric between the user channels and the fixed beamforming directions to determine the best beam for each user, $b_u^*$. NOMA-BB then clusters users that all have the same \say{best beam,} hence the name NOMA-BB. Additionally, there is a constraint on the maximum number of users, $n_{\textnormal{max}}$, not to overload certain clusters with too many users, since in practice, each digital unit has a limit on the number of parallel transmissions per time slot. If the number of users with the same BB is greater than $n_{\textnormal{max}}$, NOMA-BB simply splits them into more than one cluster, all served by the same beam but in different time slots. A similar but complementary constraint is considered in~\cite{rajasekaran_userclustering_2020} where each user had their own individual successive interference cancellation (SIC) decoding capability, \textcolor{black}{ since in practice, the NOMA users cannot perform SIC to decode an infinite number of other users’ signals due to the limited computational and energy and memory resources of the UEs.} The goal of the case study is to see if NOMA clustering can be done using only the visual feedback from the CBS and how it performs compared to approaches such as accurate CSI-based NOMA-BB. The key point is to show that CSI is not essential for determining the best beam of each user, $b_u^*$. If this $b_u^*$ can be determined through the other non-CSI-based UE feedback dimensions discussed in Section~\ref{sec:3dimesnions}, the rest of NOMA-BB can be used for NOMA clustering. 


For this case study, we expand on the \textcolor{black}{deep learning (DL)} based beam prediction scheme proposed in~\cite{alrabeiah_beamprediction_2020}, where a ResNet-18 neural network pre-trained on the popular ImageNet2012 dataset is customized for the purpose of beam prediction\footnote{As future works, we plan to investigate the performance of the system along with more complex and realistic datasets, such as ViWi version 2 (https://www.viwi-dataset.net) with multi-user scenarios, or DeepSense 6G (https://deepsense6g.net).}. \textcolor{black}{ Typical DL neural networks for image classification are skilled at classifying images into the appropriate class with sufficient training examples of images that belong to the different classes.} In~\cite{alrabeiah_beamprediction_2020}, it is shown that with a pre-defined BF codebook like in our problem, learning beam prediction from the RGB images degenerates to an image classification task where the goal of the system is to identify to which sector a user belongs. In other words, since the set of candidate beam vectors divides the scene (spatial dimensions) into multiple sectors, and single-user images are used to train the neural network, the image classification DL algorithm identifies the sectors to which a user belongs. Thus, the algorithm finds the users \say{best-beam direction} using the image captured by the CBS only. However, the images used to train the ResNet-18 neural network are single-user images (i.e., they contain only one user per image).


\begin{algorithm}[!t] 
\SetAlgoLined
\caption{Proposed DL-based NOMA clustering algorithm using the images captured by the CBS.} 
\label{fig:Algorithm}
\justify
\textbf{Stage 1 (Training)}: \texttt{\\}
1. CBS collects CSI from users and images from the scene. \texttt{\\}
2. CBS uses CSI to generate training data for the ResNet-18 beam prediction algorithm. \texttt{\\}
3. NOMA clustering is done using CSI at this stage. \texttt{\\}
\textbf{Stage 2 (Execution)}: \texttt{\\}
1. CBS uses UE's location input and the trained ResNet-18 model to pick the best beam, $b_u^*$, for each user $u$. \texttt{\\}
2. NOMA-BB: \texttt{\\}
\For{(beam-$b$ : $B_c$)}{ \texttt{\\} A. Group all $n$ users that picked $b_u^*= b$ into a cluster, to be served by beam-$b$. \texttt{\\}
B. \textbf{If} ($n=0$), do not form a cluster to be served by beam-$b$. \texttt{\\}
C. \textbf{If} ($n > n_\textnormal{max}$), split the $n$ users into $\ceil[\big]{\frac{n}{n_\textnormal{max}}}$ clusters, all served by beam-$b$.}

\noindent \textbf{Stage 3 (Validation)}: \texttt{\\}
1. Collect CSI of random users and validate against best beam, $b_u^*$, predicted by ResNet-18 model. \texttt{\\}
2. If error threshold reached, go back to Stage 1. \texttt{\\}
Note: Stage 2 and Stage 3 run in parallel.

\end{algorithm}


\begin{figure}[!t]
    \centering
    \vspace*{-.02in}
    \subfloat[DL model trained with 100 samples in Stage 1.]{
        \hspace*{-.1in}
        \includegraphics[scale=0.56]{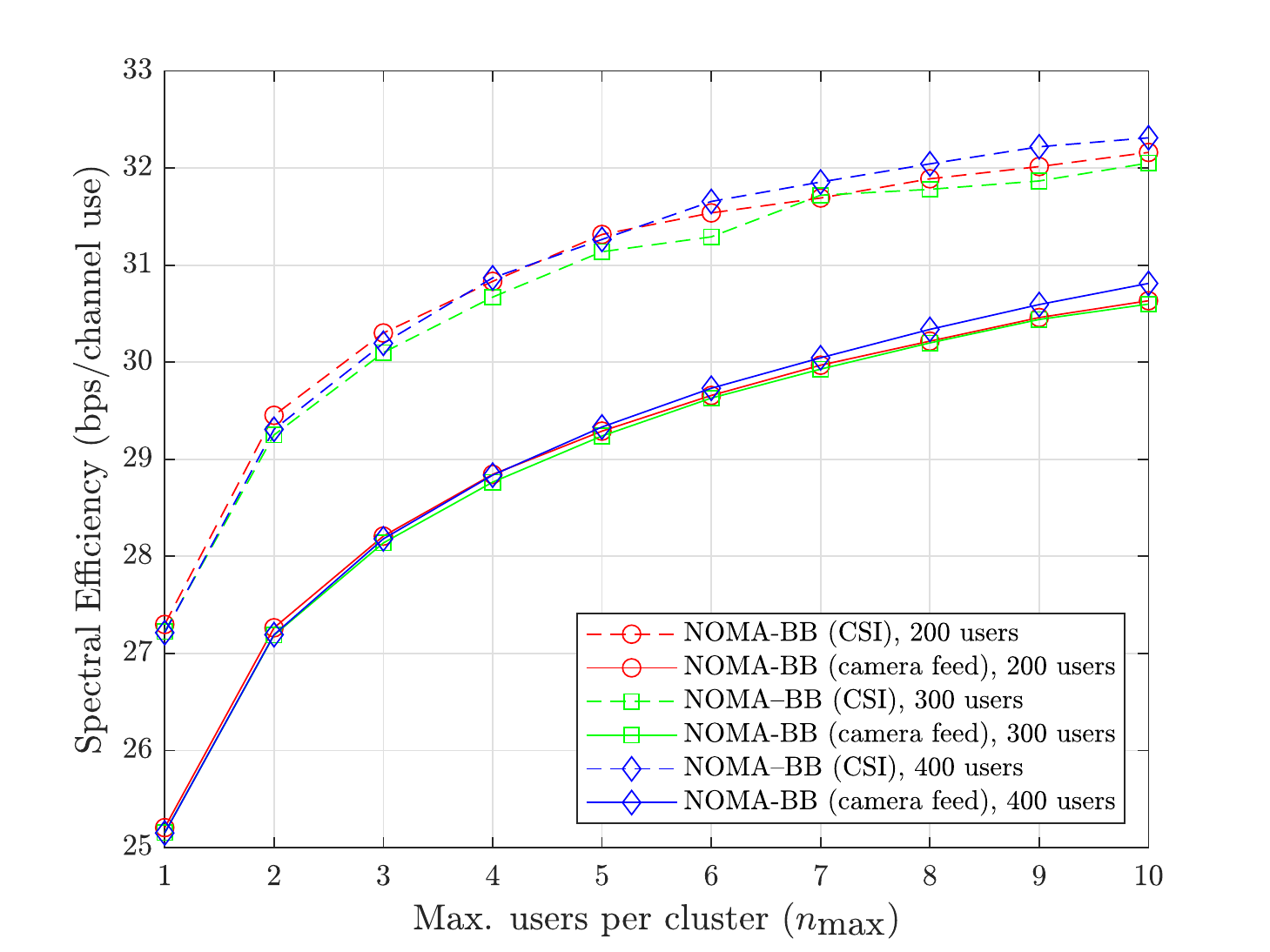}
    }
    \hfill
  \subfloat[DL model trained with 500 samples in Stage 1.]{
        \hspace*{-.1in}
        \includegraphics[scale=0.56]{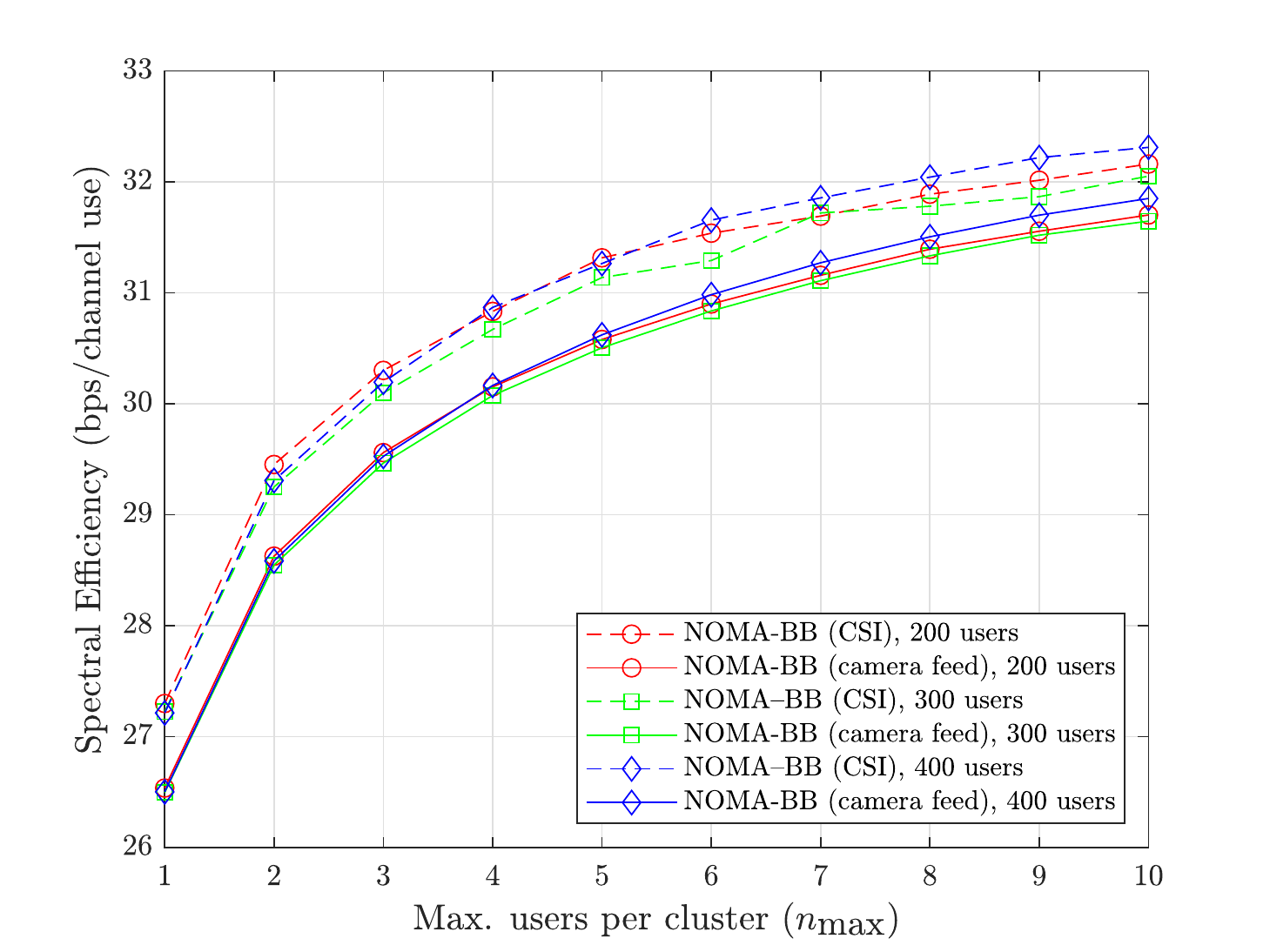}
    }
    \caption{Simulation results highlighting the close performance of the proposed DL approach using camera and location feed (non-CSI feedback) with the clustering schemes that use CSI feedback.}
\label{fig:simulation}
\vspace{-2.0em}
\end{figure}

This concept can be extended to NOMA clustering problems where a CBS will capture images with hundreds of users that it needs to serve. As described above, an algorithm like NOMA-BB can be adapted to make the user clustering decisions based on the \say{best-beam} prediction from a neural network. Fig.~\ref{fig:ClassRoom} shows what this deployment would look like in a classroom, where a CBS captures images of all users. Unlike in~\cite{alrabeiah_beamprediction_2020} that assumes single-user images, in our case and in reality, we will have multiple users in an image that are to be served by the CBS, particularly in NOMA use-cases where the goal is to serve a large number of users. However, the location information can be used to break down these multi-user images into single-user images. We just need to learn the best beam from each user's perspective, and this can be done with training examples from any user in the scene. This is because in a single-path LOS scenario with no obstacles, and where all users have equal channel gain, the best beam learned for user A at location (X,Y) implies that the best beam for user B at location (X,Y) is also the same. In more complex channel models, we need to adapt the model further, as discussed in the future research directions in Section~\ref{sec:future}.




We divide the entire algorithm into three distinct stages, as shown in Algorithm~\ref{fig:Algorithm}. The first stage involves the training phase where the CBS learns how to make \say{best beam} predictions for all users in an image. To achieve this, the CBS collects CSI of all users to determine each user's best beam using the cosine similarity metric like in~\cite{Ding2017BF_mmWaveNOMA}, and that is provided as a training sample to the DL algorithm along with a picture of the scene. During this phase, since the CBS is not yet trained to make predictions using the DL algorithm, the CSI is used to make NOMA clustering decisions to provide system continuity. Once the DL algorithm is sufficiently trained, the CBS moves to Stages 2 and 3 jointly. In Stage 2, the CBS stops collecting the users' CSI and instead starts using the beam predictions made by the trained DL algorithm to feed to NOMA-BB for the final NOMA clustering decisions. In parallel to this, Stage 3 is run where the CSI of a few users is collected in order to validate against the best beam prediction from the DL model. If a defined error threshold is reached, the algorithm reverts to Stage 1 if the environment has sufficiently changed from the originally trained model. For instance, in LoS dominated scenarios like classrooms, coffee shops, etc. where there can be many users to serve (user devices like today, but also internet-of-things (IoT) devices in the future), the environment does not change that often and so we will not have to keep reverting from Stage 3 to Stage~1. 



The performance of this proposed scheme is shown in Fig.~\ref{fig:simulation}, where we can see that the DL camera-based clustering scheme is able to achieve comparable performance in spectral efficiency to the CSI-based scheme. We can see that the performance of the camera feed approach improves as more training data is used in Stage 1 to train the ResNet-18 based DL model. The difference in performance between the CSI and camera feed schemes shown here is entirely due to the beam prediction aspects. Errors in predicting the best beam for some users mean they are placed in clusters where they do not receive the maximum possible signal-to-noise and interference ratio (SINR). It is worth mentioning that the user ordering and power allocation parts of the NOMA scheme depend on the CSI of the user, as it is most optimal to order the users for SIC decoding in the order of their channel gains. In a truly CSI-independent scheme, we would have to do this arbitrarily, which would cause a loss of performance as well. 



\section{Challenges and Future Directions} \label{sec:future}

In this section, we discuss potential challenges and future research directions for implementing DL-based NOMA clustering algorithms using the images captured by CBSs. 

\textbf{Camera coverage and cost}: The number of cameras used, the placement of these cameras, and the quality of the pictures captured by the CBS are all interesting questions that will have an impact on the NOMA performance. Of course, better quality images equal more cost, so exploring such trade-offs, particularly for low-cost small cell BSs is an important aspect for practical deployments. 

\textcolor{black}{\textbf{Frequency of camera updates and mobility aspect}: A cluster of users has to be determined by the baseband unit on the order of hundreds of milliseconds. The current camera feed might not provide updates that often and then have clustering decisions be made accordingly. Hence, the control interface between providing new camera updates, making beam predictions and accordingly NOMA clustering decisions is an important challenge to address. On the other hand, in this paper, it is assumed that user locations do not change faster than every few seconds, since in practice, how often the user location (i.e., beam-index information) changes will have a direct impact on the load of the control interface and the design of the interface.}

\textbf{Imprecise location information}: Another challenge is addressing imprecise location information. Since the system relies on the location information to identify users on the scene, other object detection techniques such as~\cite{szegedy_object_detection} can be used for user identification to complement user location~feedback. 

\textbf{Complexity of channel models and amount of data}: In multi-path settings, learning the best beam from a set of candidate beams could involve more advanced features, and thus also, require more training data. For example, multi-path poses a challenge to the neural network to learn the requisite advanced features of the channel. Additionally, users themselves can be obstacles and alter the multi-path setting and so the best beam of the users. 

\textcolor{black}{\textbf{Privacy and security}: An important concern with the use of cameras for clustering is the user privacy and security concerns that come with it. This is true for any vision-assisted scheme where the BS gets access to camera images that identify individual users. With the proposed scheme, given that the camera feed is used only for 0clustering decisions, regulations can be built-in to ensure the BS does not store any history related to the user. 
For cases where the pictures might be stored for some time for offline training of the model, the user context and identity of the users can be removed; so nothing can be traced back.}

\textcolor{black}{\textbf{Anticipatory networking with visual information}: 
To reduce operational and cost inefficiencies of the next-generation wireless networks, knowing the future user distribution in both spatial and temporal domains, i.e., forecasting the future state of the network, at various time scales is critical~\cite{bui2017survey}. Visual information collected by various cameras on the BSs can be leveraged to predict user distribution with high precision, and to optimize not only the network functions such as handover, but also, the network performance through such as cell dimensioning, cell switch-off, and load balancing.}

\section{Conclusion}

\textcolor{black}{In this paper, the potential of CBSs in mmWave-NOMA systems was explored to improve the robustness of user clustering.} We showed three different dimensions of user feedback that a CBS can exploit for user clustering, depending on the situation; CSI-based feedback or non-CSI-based feedback, including the UE location and camera feed of the CBS. Exploiting the advances in deep learning, the NOMA clustering problem can be addressed using the images captured by CBSs, without consuming extra RF resources. Through a case study, we showed that such an approach achieves comparable performance to a CSI-based approach\textcolor{black}{, and user clustering can continue to function without much performance loss even in the scenarios where CSI is severely outdated or not available at all.} Lastly, open challenges and future research directions in this area were discussed. 

\section*{Acknowledgment}
The work of Omar Maraqa and Saad Al-Ahmadi was supported by the interdisciplinary research center for communication systems and sensing (IRC-CSS), King Fahd University of Petroleum and Minerals, under Grant number INCS2107.

\bibliographystyle{adityaIEEEtran}
\bibliography{IEEEabrv,main}

\end{document}